\documentclass[%
reprint,
amsmath,amssymb,
aps,
superscriptaddress,
prl
]{revtex4-1}

\usepackage{graphicx}
\usepackage{dcolumn}
\usepackage{bm}
\usepackage{hyperref}
\usepackage{siunitx}

\begin{document}

\preprint{APS/123-QED}

\title{Superconductivity in Li-intercalated 1T-SnSe$_2$ driven by electric-field gating}

\author{Yanpeng Song}
\affiliation{Beijing National Laboratory for Condensed Matter Physics, Institute of Physics, Chinese Academy of Sciences, Beijing 100190, China}
\affiliation{University of Chinese Academy of Sciences, Beijing 100049, China}

\author{Xiaowei Liang}%
\affiliation{Center for High Pressure Science, State Key Laboratory of Metastable Materials Science and Technology, Yanshan University, Qinhuangdao 066004, China}%

\author{Jiangang Guo}
\email{jgguo@iphy.ac.cn}
\affiliation{Beijing National Laboratory for Condensed Matter Physics, Institute of Physics, Chinese Academy of Sciences, Beijing 100190, China}
\affiliation{Songshan Lake Materials Laboratory, Dongguan, Guangdong 523808, China}

\author{Jun Deng}
\affiliation{Beijing National Laboratory for Condensed Matter Physics, Institute of Physics, Chinese Academy of Sciences, Beijing 100190, China}
\affiliation{University of Chinese Academy of Sciences, Beijing 100049, China}

\author{Guoying Gao}
\affiliation{Center for High Pressure Science, State Key Laboratory of Metastable Materials Science and Technology, Yanshan University, Qinhuangdao 066004, China}%

\author{Xiaolong Chen}
\email{xlchen@iphy.ac.cn}
\affiliation{Beijing National Laboratory for Condensed Matter Physics, Institute of Physics, Chinese Academy of Sciences, Beijing 100190, China}
\affiliation{University of Chinese Academy of Sciences, Beijing 100049, China}
\affiliation{Songshan Lake Materials Laboratory, Dongguan, Guangdong 523808, China}

\date{\today}

\begin{abstract}
Creating carrier reservoirs in layered compounds can effectively tune the carrier density, which often induces a variety of emergent properties. Based on solid-ion-conductor gating technique, we successfully induce superconductivity of 4.8 K in ultrathin Li-intercalated SnSe$_2$ samples. The Li$^+$ ions are driven in between interspacing of SnSe$_2$ layers and form a single reservoir layer to provide electrons. In addition, a dome-like \emph{T}$_c$ is found through substituting of S for Se, where the optimal \emph{T}$_{c}$ is 6.2 K for SnSe$_{1.8}$S$_{0.2}$. Density functional theory calculations confirm that the intercalated LiSnSe$_2$ is thermodynamically favorable, where the intercalation of Li expands the interlayer spacing by 10\% and increases the carrier density by two orders of magnitude. Meanwhile the calculated results reveal that the enhanced electron-phonon interaction due to softened phonon determines the occurrence of superconductivity. Our results demonstrate that this strategy is very effective to explore superconductors in layered materials with narrow bandgap.

\end{abstract}

\pacs{Valid PACS appear here}

\maketitle
\section{\uppercase\expandafter{\romannumeral1}. Introduction}

Carrier tuning is essential for inducing many emergent properties such as superconductivity (SC)\cite{RN1}, stripe phase\cite{RN2}, pseudogap\cite{RN3}, charge density wave\cite{RN4} and nematic phase\cite{RN5} in cuprate and iron-based superconductors. The ways of hole doping (Ba$^{2+}$ for La$^{3+}$) and electron doping (F$^-$ for O$^{2-}$) in La$_2$O$_2$ layer successfully induce SC at 38 K and 26 K in La$_2$CuO$_4$ and LaFeAsO, respectively\cite{RN6,RN7}. Other route like exertion of chemical or physical pressure plays a similar role as chemical doping does, as manifested by induced SC in AeFe$_2$As$_2$ (Ae=Ca, Sr, Ba)\cite{RN8,RN9,RN10,RN11}. The emergence of SC upon such tuning routes is related to change the curvature of energy band and geometry of Fermi surface.

For those compounds without carrier-supplying layer, creating a spacer layer in a Van der Waals gap is an effective strategy for inducing SC. For example, inserting a spacer layer of K$^+$ and Tl$^+$/Rb$^+$/Cs$^+$ in between FeSe layers can form A$_x$Fe$_{2-y}$Se$_2$ phase (A=K, Tl, Rb, Cs) and enhance the superconducting critical temperature (\emph{T}$_c$) in bulky FeSe from 8 K to 31 K\cite{RN12,RN13,RN14,RN15}. It is worthy to note that the intercalation of Na$_x$(NH$_3$)$_y$ or [Li$_{0.8}$Fe$_{0.2}$OH]$^{+\sigma}$ layer into FeSe layer further increases \emph{T}$_c$ to 41-45 K, although it is not stable at elevated temperatures\cite{RN16,RN17}. Hence, low-temperature synthesis routes like hydrothermal and solvothermal methods are conducive to obtain those superconductors\cite{RN18,RN19}. Recently, a novel intercalated method through electric-field gating has been used to induce the highest \emph{T}$_c$ (48 K) in FeSe-based superconductors at ambient pressure \cite{RN20,RN21,RN22}. The back-gate applied on solid-ion-conductor of Li$_2$Al$_2$SiP$_2$TiO$_{13}$ can drive the cation (Li$^+$) into FeSe precursor, and thus enhance carrier concentration and \emph{T}$_c$. It provides a relatively clean process of carrier tuning without drastically altering the crystal structure of parent compound. One thing should be noted is that this method differs from that of ionic-liquid gating, where no ions migrate into the precursor and thus the SC is confined within surface layers (thickness of $\sim$ 1 nm)\cite{RN23,RN24}.

SnSe$_2$ is an intrinsic semiconductor with a bandgap of 1.0 eV at room temperature\cite{RN25}. Intercalation of Co($\eta$-C$_5$H$_5$)$_2$ molecule between SnSe$_2$ layers can induce SC at  \emph{T}$_c$=5 K\cite{RN26,RN27}. Besides, two-dimensional SC has been found in the interface of SnSe$_2$/ionic-liquid and SnSe$_2$/graphene\cite{RN28,RN29}. Here, we choose ultrathin 1T-phase SnSe$_2$ with thickness $\sim$ 13 nm as a host to testify the carrier-tuning effect on transport properties through electric-field gating. We find that a transition from semiconductor to superconductor for SnSe$_2$ due to Li-intercalation driven by electric-field. The highest \emph{T}$_c$, 6.2 K, occurs as combining Li-intercalation and partial chemical substitution of S for Se. Theoretical calculations support that the Li ions might occupy interlayer position to form a carrier reservoir. Within the framework of Bardeen-Cooper-Schrieffer (BCS) theory, our calculations show that electron-phonon coupling (EPC) constant ($\lambda$) is 0.97 and the phonon-mediated mechanism is responsible for SC.

\section{\uppercase\expandafter{\romannumeral2}. EXPERIMENTAL}
	\subsection{A. Experimental details}
Single crystal of SnSe$_2$ was grown by chemical vapor transport method in a two-zone furnace (\emph{T}$_1$=1000 K and \emph{T}$_2$=800 K) with iodine as transport agent. Selected area electron diffraction (SAED) was carried out on a transmission electron microscopy (JEM-2100 Plus). Standard mechanical-exfoliation methods with Scotch tapes are used to obtain thin flakes of SnSe$_2$. A step-by-step process is illustrated in Fig. 1. We exfoliated thin flakes of SnSe$_2$ from bulk crystal by Scotch tapes. The thickness of flake was characterized by atomic force microscope (AFM) (Bruker Multimode 8). Typical samples of 10 $\sim$ 30 nm were transferred on solid-ion conductor by a piece of polydimethylsiloxane (PDMS). Au film of 50 nm was pre-deposited on the opposite surface of solid-ion conductor through electron beam evaporation. The conductor was placed on a Au-coated SiO$_2$/Si substrate. We then deposited Ti/Au or Cr/Ag (5/50 nm) metal electrodes with standard four-terminal and Hall-terminal patterns on SnSe$_2$ surface. The whole device was then quickly loaded into a refrigerator to avoid absorption of water or oxygen from the air. It is known that Li-ion conductivity of solid-ion-conductor Li$_2$Al$_2$SiP$_2$TiO$_{13}$ (HF-Kejing) is 10$^{-4}$ S cm$^{-1}$ as T$>$160 K. Thus, we slowly increased T to 160 K in case of overcharging. A lock-in amplifier (Stanford Research 830) was used to measure the transport properties. The gate voltage, V$_G$=4.5 V, was supplied by a Keithley-2400 source meter. The resistance of SnSe$_2$ started to drop upon applying electric-field, meaning that Li ions of Li$_2$Al$_2$SiP$_2$TiO$_{13}$ are driven into SnSe$_2$, see Fig. S1. Meanwhile, the intercalation of Li ions is also confirmed by the I$_{DS}$-V$_G$ curve as shown in Fig. S2, where the positive voltage can effectively reduce resistance and increase I$_{DS}$.

\subsection{B. Computational details}
Our calculations were based on density functional theory (DFT) by using plane-wave pseudopotential method as implemented in the Vienna Ab initio Simulation Package\cite{RN30}. The electron–ion interactions were described by the projector-augmented wave potentials with 5\emph{s}$^2$5\emph{p}$^2$, 4\emph{s}$^2$4\emph{p}$^4$ and 2\emph{s}$^1$ configurations treated as valence electrons for Sn, Se and Li, respectively. The exchange-correlation function was described using Perdew-Burke-Ernzerhof\cite{RN31} of generalized gradient approximation. A kinetic cutoff energy of 600 eV and corresponding Monkhorst-Pack (MP)\cite{RN32} \emph{k}-point meshes for different structures were adopted to ensure that the enthalpy converges to 1 meV/atom. The $\lambda$ of LiSnSe$_2$ was calculated within the framework of linear response theory through the Quantum-ESPRESSO code\cite{RN33} , where ultrasoft pseudopotentials for Sn, Se and Li with a kinetic energy cutoff of 70 Ry were employed. We adopted a 9$\times$9$\times$5 \emph{q}-point mesh and 36$\times$36$\times$20 MP \emph{k}-point mesh in the first Brillouin zone (BZ) for the EPC calculations.

\begin{figure}
	\includegraphics{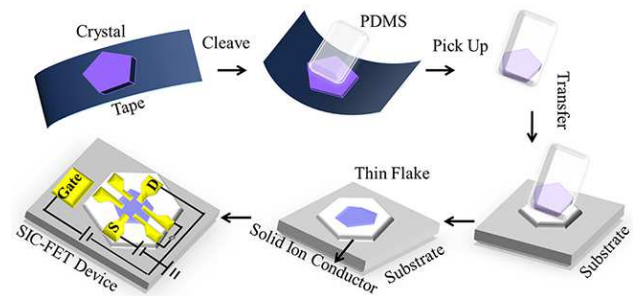}
	\caption{\label{fig1}Step-by-step process for fabricating SnSe$_2$/solid-ion-conductor device. Yellow electrode patterns are deposited on the surface of SnSe$_2$ and solid-ion-conductor, respectively.}
\end{figure}

\section{\uppercase\expandafter{\romannumeral3}. RESULTS AND DISCUSSION}
SnSe$_2$ only has a 1T-type crystal structure with \emph{a}=3.807 \AA\ and \emph{c}= 6.128 \AA\ (space group: \emph{P}$\bar{3}$m1), differing from the polymorphism of most transition metal dichalcogenides. Figure 2 (a-c) show crystal structures of SnSe$_2$ and slightly-distorted SnSe$_6$ octahedron. Each Sn is surrounded by six Se with a Sn-Se bondlength of 2.88 \AA\  and a Se-Sn-Se angle of \ang{93}. The interlayer spacing is as large as 3.07 \AA, implying that the interaction is weak Van de Waals force. In Fig. 2(d), the SAED along [001] zone axis confirms that the as-grown single crystal has high crystallinity. From the AFM image, we can find that the exfoliated SnSe$_2$ is $\sim$ 13 nm thick, i. e. 20 unit cell, as shown in Fig. 2(e) and (f). The optical image of sample and four electrodes are shown in Fig. 2(g), and the transverse size of sample is $\sim$50 $\mu$m.

\begin{figure}
	\includegraphics{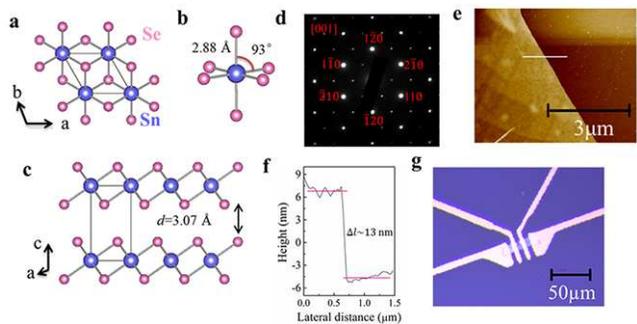}
	\caption{\label{fig2}(a-c) Crystal structure of SnSe$_2$ and SnSe$_6$ octahedron. The Sn-Se bondlength, Se-Sn-Se angle and interlayer spacing are labeled. (d) Selected area electron diffraction (SAED) of single crystal of SnSe$_2$ along [001] zone axis. (e) AFM topographic image of SnSe$_2$ thin flake. (f) Cross-sectional profile of SnSe$_2$ along the white line in \textbf{e}. The thickness ($\Delta$\emph{l}) is 13 nm. (g) Optical image of four electrodes on SnSe$_2$. }
\end{figure}

The temperature-dependent sheet electrical resistance (\emph{R}$_s$) of SnSe$_2$ is plotted in Fig. 3(a). The curve shows overall semiconducting character. The data in intermediate temperature range can be fitted by equation, \emph{R}(T)=\emph{R}(0)exp[\emph{E}$_g$/\emph{k}$_B$T], where \emph{E}$_g$ is thermal-active gap, \emph{k}$_B$ the Boltzmann constant. Fitting the curve yields an \emph{E}$_g$ = 5 meV, which is much smaller than the bandgap of bulk SnSe$_2$ ($\sim$ 1.0 eV). The Hall coefficient (\emph{R}$_H$) and carrier concentration (\emph{n}) of SnSe$_2$ were calculated from magnetic-field-dependent Hall resistance (\emph{R}$_{xy}$). Based on the linear relationship between \emph{R}$_{xy}$ and magnetic field (B) up to 7 T, see Fig. 3(b), we can fit the slop and obtain the \emph{R}$_H$=\emph{R}$_{xy}$/B. The \emph{R}$_H$ is negative in the measured temperature range. Temperature-dependent \emph{R}$_H$ and \emph{n} are plotted in Fig. 3(c). The \emph{R}$_H$ gradually decreases from -3 m$^2$/C to -10 m$^2$/C with decreasing temperature. Thus, the \emph{n}, calculated from single-band model \emph{n}=1/e\emph{R}$_H$, decreases from 2.2$\times$10$^{14}$ cm$^{-2}$ to 6$\times$10$^{13}$ cm$^{-2}$ from 300 K to 80 K. This value is consistent with relatively high carrier concentration in few-layer SnSe$_2$\cite{RN34}, while it is higher than 10$^{12}$ -10$^{13}$ cm$^{-2}$ of few-layer MoS$_2$ and MoSe$_2$\cite{RN35,RN36}.

\begin{figure}
	\includegraphics{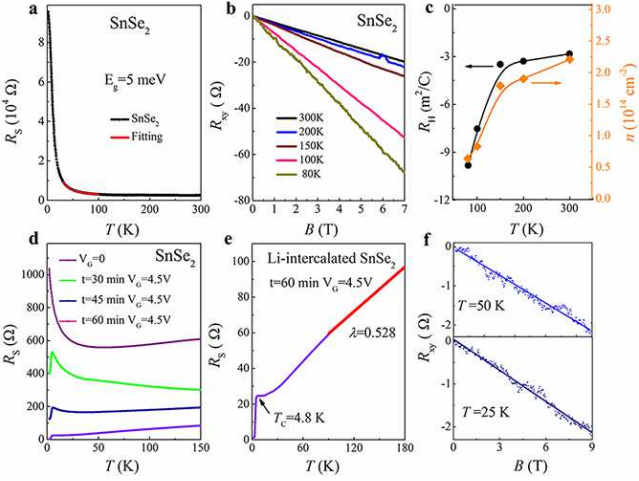}
	\caption{\label{fig3}(a) Temperature-dependent \emph{R}$_{s}$ of SnSe$_2$ thin flake without applying gating voltage. (b) Magnetic-field dependence of Hall resistance (\emph{R}$_{xy}$) measured at 80 K, 100 K, 150 K, 200 K and 300 K. (c) Temperature dependence of the Hall coefficient (\emph{R}$_{H}$) and 2D carrier concentration (\emph{n}) of SnSe$_2$. (d) Temperature-dependent \emph{R}$_{H}$ of Li-intercalated SnSe$_2$ between 2 K and 150 K under V$_G$= 4.5 V with different gating time (t). (e) Superconducting transition occurs at \emph{T}$_{c}$ = 4.8 K. Red line is fitting curve from 90 K to 180 K. (f) Magnetic field dependence of \emph{R}$_{xy}$ of Li-intercalated SnSe$_2$ measured at 25 K and 50 K, respectively.  }
\end{figure}

To verify the effect of electric-field gating on SC, we measure the transport properties of SnSe$_2$ down to 2 K under a positive V$_G$ = 4.5 eV. Fig. 3(d) and (e) are the variation of the \emph{R}$_s$ against temperature. We found that the SnSe$_2$ changes into metal as increasing gating-time (t), meaning that Li$^+$ driven by electric-field enters into SnSe$_2$ so that Li-intercalation increases the carrier concentration. Meanwhile, a slight drop of resistance shows up at 4.8 K, and this transition gradually enhance as increasing t. From Fig. 3(e), one can find that a superconducting transition occurs at \emph{T}$_{c}^{onset}$ = 4.8 K and \emph{T}$_{c}^{zero}$ = 3.5 K, which are comparable to those \emph{T}$_{c}$ in SnSe$_2$-based superconductors. Here, we use the electron-phonon scattering model to analyze the $\lambda$ of Li-intercalated SnSe$_2$. In the high-temperature limit (T/$\Theta_D$ $\gg$ 1, where  $\Theta_D$ is the Debye temperature), the resistivity can be expressed as\cite{RN37,RN38}
\begin{eqnarray}
\rho\approx\rho_0+\lambda\frac{2\pi mk_B}{\hbar e^2n_0}T
\end{eqnarray}
where $\rho_{0}$ is temperature-independent resistivity due to electron-impurity scattering, $\lambda$ EPC constant, \emph{m} effective mass, \emph{k}$_{B}$ the Boltzmann constant and \emph{n}$_{0}$ volume carrier density. For SnSe$_2$ sample with thickness $\Delta$\emph{l}, the \emph{R}$_{s}$ and \emph{n} are \emph{R}$_{s}$ =$\rho_{0}$/$\Delta$\emph{l} and \emph{n} = \emph{n}$_{0}$$\times$$\Delta$\emph{l}, respectively. Therefore,
\begin{eqnarray}
	R_s\approx R_0+\lambda\frac{2\pi m k_B  }{\hbar e^2n }T=R_0+\lambda\frac{2\pi\hbar k_B}{\varepsilon_0\Delta\emph{l}(\hbar\omega_{p})^2}T
\end{eqnarray}

where the bulk plasma frequency ($\hbar\omega_{p}$) for SnSe$_2$ is 1.01$\times$10$^5$ cm$^{-1}$\cite{RN39}, $\varepsilon_0$ the vacuum permittivity. Fitting temperature dependent \emph{R}$_{s}$ from 90 K to 180 K yields $\lambda$ is 0.53. In addition, the Hall resistance \emph{R}$_{xy}$ of Li-intercalated SnSe$_2$ is measured to estimate the carrier concentration. The magnetic-filed dependent \emph{R}$_{xy}$ at 25 K and 50 K are plotted in Fig. 3(f). We extract \emph{R}$_{H}$ from the linear fitting of magnetic-filed-dependent \emph{R}$_{xy}$ and calculate the \emph{n}=1/e\emph{R}$_{H}$. It turns out that application of gate voltage, V$_G$ = 4.5 V, increases the \emph{n} as high as $\sim$2.58$\times$10$^{15}$ cm$^{-2}$ at T = 25 K.

Next, we investigate the superconductivity of electric-field gated SnSe$_{2-x}$S$_x$ (\emph{x}=0.2, 0.4, 0.5) samples and present the data in Fig. 4. Similar superconducting transitions were also observed in SnSe$_{2-x}$S$_x$  (\emph{x}= 0.2, 0.4, 0.5) under the same V$_G$ = 4.5 eV, as shown in Fig. 4(a). Fig. 4(b) shows that the resistivity gradually decrease with increasing gating time, and the optimal transition can be reached as t= 60 min.  The temperature-dependent \emph{R}$_s$ of SnSe$_{1.8}$S$_{0.2}$  under varying perpendicular magnetic fields  is shown in Fig. 4(c). It can be seen that the superconducting transition gradually broadens and the \emph{T}$_{c}$ shifts to lower temperatures as the magnetic field increases. Meanwhile, as a criteria of 90\% of the normal state resistance, we plotted the temperature dependence of the field \emph{H}$_{c2}$ for Li-intercalated SnSe$_{1.8}$S$_{0.2}$ in the inset of Fig. 4(c).
These data can be fitted by the Ginzburg-Landau equation $\emph{H}_{c2}$ = $\emph{H}_{c2}(0)$[1-($\frac{T}{T_c})^2$] and the obtained $\emph{H}_{c2}(0)$ = 0.39 T. Using the equation of $\emph{H}_{c2}(0)$ = $\frac{\Phi_0}{2\pi\xi_{GL}(0)^2}$, where $\Phi_0$ is the magnetic flux quantum (2.07 × $\times$10$^{-15}$ Wb), we obtained the  ${\xi_{GL}(0)}$ = 29 nm. A superconducting phase diagram of \emph{T}$_{c}$ versus S content is plotted in Fig. 4 (d). In the  SnSe$_{2-x}$S$_x$  sample (x$\le$0.5),
an interesting point is that the \emph{T}$_c^{onset}$ firstly increases to 6.2 K at \emph{x}=0.2, and then it begins to decrease as \emph{x}$>$0.2. It is known that  the S-substitution can induce contraction of lattice constants and Sn-Se/S bondlength, which generates the effect of chemical pressure without introducing extra electrons\cite{RN40}. As for the S-substitution effect on superconductivity, we think that it is similar with the observations in FeSe$_{1-x}$S$_x$, in which the \emph{T}$_c^{onset}$ firstly increases and then decreases as increasing S, and the maximal \emph{T}$_c^{onset}$ emerges at 20\% S-substitution. This enhancement \emph{T}$_c^{onset}$ should be related to the optimized Fe-Se bondlength and geometry of FeSe$_4$ tetrahedra\cite{RN41}. Another possibility is that the substitution of S for Se results in a greater orbital overlap, increasing the bandwidth and driving the system more itinerant\cite{RN42}. In the S-rich SnSe$_{2-x}$S$_x$  sample (x$>$0.5), the \emph{R}$_s$ is too large to be measured\cite{RN43}. The electric-field gating is hard to metallize this sample and induce superconductivity. We notice that, in all samples, the \emph{T}$_c^{onset}$ does not increase as increasing t. It implies that there is only one superconducting phase, which is different from multiple phases in intercalated FeSe-based superconductors\cite{RN22,RN44,RN45}.

\begin{figure}
	\includegraphics{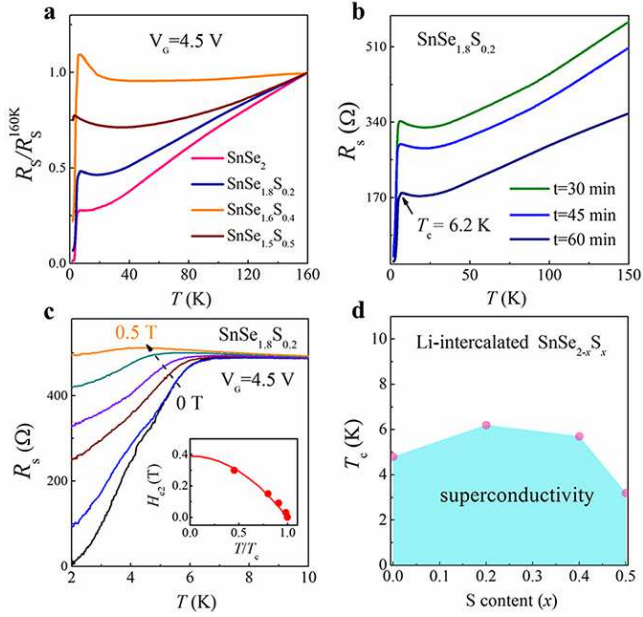}
	\caption{\label{fig4}(a) Superconducting transitions were observed in all SnSe$_{2-x}$S$_x$ (x=0.2, 0.4, 0.5) samples under V$_G$= 4.5 V. (b) Temperature-dependent \emph{R}$_{s}$ between 2 K and 150 K of SnSe$_{1.8}$S$_{0.2}$ with different gating time. (c) Temperature dependence of the magnetoresistance of SnSe$_{1.8}$S$_{0.2}$. The inset shows the temperature dependence of \emph{H}$_{c2}$  for  SnSe$_{1.8}$S$_{0.2}$. The red line denotes the fitting curve obtained from the Ginzburg-Landau model. (d) Superconducting phase diagram of SnSe$_{2-x}$S$_x$. The \emph{T}$_{c}$ firstly increases to 6.2 K at x=0.2, and then it decreases as x$>$0.2.  }
\end{figure}

To better understand the effect of Li-intercalation on the structure and SC of SnSe$_2$, the DFT calculations were performed. Here, four stoichiometries of Li-intercalated phases, LiSnSe$_2$, LiSn$_2$Se$_4$, LiSn$_3$Se$_6$ and LiSn$_4$Se$_8$, were adopted to simulate different Li concentrations. Their crystal structures were constructed based on the \emph{P}$\bar{3}$m1 space group, where one Li atom occupies one octahedral interstice surrounding by six Se atoms (Wyckoff position of \emph{P}$\bar{3}$m1:1b) in 1$\times$1$\times$1, 1$\times$1$\times$2, 1$\times$1$\times$3 and 2$\times$2$\times$1 supercells of LiSnSe$_2$, respectively. We first investigate the phase stability of Li-intercalated SnSe$_2$ by calculating their formation enthalpies relative to Li and binary SnSe$_2$, where the most stable structures for Li and SnSe$_2$ at ambient pressure were adopted\cite{RN46}. As shown in Fig. 5(a), all the stoichiometries have negative formation enthalpies relative to Li and SnSe$_2$, suggesting that the intercalation of Li is thermodynamically favored. With increasing Li content, the formation enthalpies monotonously decreases and LiSnSe$_2$ achieves the lowest one, -0.42 eV/atom, indicating that it is the most stable compound. LiSn$_2$Se$_4$, LiSn$_3$Se$_6$ and LiSn$_4$Se$_8$ were calculated to be less stable with respect to LiSnSe$_2$. The obtained crystal structure of LiSnSe$_2$ is shown in Fig. 5(b), where the \emph{a}= 3.955 \AA\ and \emph{c}= 6.751 \AA. The Li ions intercalate into interlayer of SnSe$_2$ and form a carrier-supply reservoir. The interlayer spacing is expanded to 3.38 \AA\ after Li interaction, but the Sn-Se bondlength is still 2.88 \AA. The calculated electronic band structure and projected density of states (DOS) for LiSnSe$_2$ are shown in Fig. 5(c). Comparing with the band structure of semiconducting SnSe$_2$ in  Fig. S3 and Ref.\cite{RN47}, doping of one Li$^+$ seems to upshift the Fermi level (\emph{E}$_F$) by $\sim$ 1 eV. One can see that a single band crosses the \emph{E}$_F$ from -1 eV to 1 eV, indicative of metallic behavior, which is consistent with the experimental result. The electronic DOS at \emph{E}$_F$  is 2.1 states/eV, which is dominated by Sn 5\emph{s} and Se 4\emph{p} orbitals with strong hybridization. The contribution of Se 4\emph{p} is a little larger than that of Sn 5\emph{s}. From the calculated Fermi surface of LiSnSe$_2$, shown in Fig. 5(d) and (e), we can see that a three-dimensional-like characteristic in electronic structure appears. There is a small electron pocket at \emph{$\Gamma$} point of the first Brillouin zone.

\begin{figure}
	\includegraphics{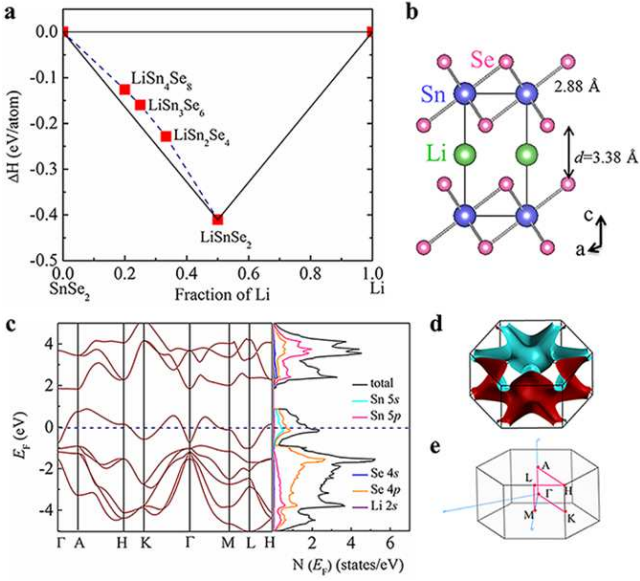}
	\caption{\label{fig5}(a) Relative enthalpies of formation per atom of different Li-SnSe$_2$ compositions. (b) Predicted structure of LiSnSe$_2$. (c) Band structure and projected density of states for LiSnSe$_2$. (d-e) Fermi surface and the first Brillouin zone for LiSnSe$_2$.  }
\end{figure}

The calculated phonon spectrum for LiSnSe$_2$ is plotted in the left panel of Fig. 6(a). There is no imaginary frequency in the full spectrum, indicating that it is dynamically stable. The corresponding projected phonon DOS (PDOS) is shown in the right panel of Fig. 6(a). It can be seen that the low frequency bands below 4 THz are dominated by Sn and Se states, while the high frequency modes above 5 THz mainly comes from Li vibrations. In comparison with the phonon spectrum and PDOS for SnSe$_2$, see Fig. S3(b and d), the intercalation of Li drives the partial phonon modes below 4 THz move to low-frequency range and enhances the peak at 2 THz. It can induce phonon softening, which enhance the electron-phonon coupling and facilitate superconductivity. To explore the SC of LiSnSe$_2$, the Eliashberg spectral function $\alpha^2$F($\omega$) and the $\lambda$ are calculated as shown in Fig. 6(b). The spectra function  $\alpha^2$F($\omega$) is calculated in terms of the phonon linewidth $\gamma_{qj}$ due to electron-phonon interaction

\begin{eqnarray}
\alpha^2F(\omega)=\frac{1}{2\pi N(E_F)}\Sigma_{qj}\frac{\gamma_{qj}}{\hbar \omega_{qj}}\delta(\omega-\omega_{qj})
\end{eqnarray}
where \emph{N}(\emph{E}$_F$) is the electronic DOS per atom and spin at the \emph{E}$_F$. The linewidth of a phonon mode, $\gamma_{qj}$, can be expressed as

\begin{eqnarray}
\gamma_{qj}=2\pi\omega_{qj}\Sigma_{knm}|g_{kn,k+qm}^j|^2\delta(\varepsilon_{kn})\delta(\varepsilon_{k+qm})
\end{eqnarray}
where the sum is over the first Brillouin zone, and $\varepsilon_{kn}$ are the energies of bands measured with respect to the \emph{E}$_F$ at point k. The \emph{g}$_{kn,k+qm}^j$ is electron-phonon matrix element. The $\lambda$ is thus defined as integration of the spectra function in the first Brillouin zone as the formula $\lambda$=2$\int_{0}^{\infty}$$\frac{\alpha^2F(\omega)}{\omega}d\omega$. 
Besides the modes at about 4 THz at \emph{$\gamma_{qj}$} point, the modes between 1-2 THz at \emph{H} point and along \emph{K}-\emph{$\Gamma$} direction soften, which would enhance the $\gamma_{qj}$. Accordingly, it is clearly seen that a pronounced peak shows up at 1.5 THz in the $\alpha^2$F($\omega$)/$\omega$ curve. The calculated $\lambda$ for LiSnSe$_2$ is 0.97, which mainly comes from the vibrations of Sn and Se atoms below 2 THz. Therefore, the main effect of Li-intercalation is believed to not only provide carriers for SnSe$_2$ layers but also enhance the electron-phonon coupling. 

\begin{figure}
	\includegraphics{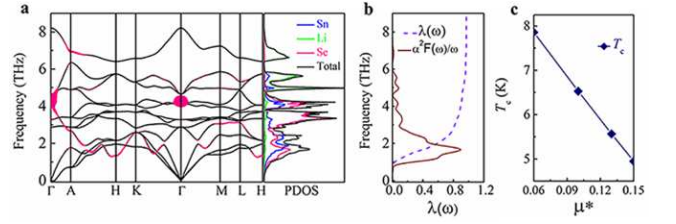}
	\caption{\label{fig6}(a) Calculated phonon spectra and PDOS. Magenta solid circles show the phonon linewidth with a radius proportional to the EPC strength. (b) Eliashberg spectral function $\alpha^2$F($\omega$)/$\omega$ and EPC constant ($\lambda$) as a function of frequency. (c) Coulomb repulsion constant $\mu^*$  dependence of \emph{T}$_c$.}
\end{figure}

We use the Allen-Dynes formula\cite{RN48}
\begin{eqnarray}
\omega_{ln}=exp[(\frac{2}{\lambda})\int_{0}^{\infty}\frac{d\omega}{\omega}\alpha^2F(\omega)log\omega]\\
T_c=\frac{\omega_{ln}}{1.2}exp[-\frac{1.04(1+\lambda)}{\lambda-\mu^*(1+0.62\lambda)}]
\end{eqnarray}
to accurately evaluate the \emph{T}$_c$, where $\omega_{ln}$ is the logarithmic average frequency, $\mu^*$ is the Coulomb repulsion constant.  The $\omega_{ln}$ is calculated to be 98.2 K, which is significantly softened by 40\% compared with the Debye temperature of bulk SnSe$_2$ (140$\pm$2 K)\cite{RN49}. The $\mu^*$  dependence of \emph{T}$_c$ is plotted in Fig. 6(c). The calculated  \emph{T}$_c$ linearly decreases from 7.8 K to 4.8 K as increasing $\mu^*$. Taking the common value of the $\mu^*$ as 0.13, the theoretical  \emph{T}$_c$ of LiSnSe$_2$ is 5.5 K. Notably, the theoretical \emph{T}$_c$ is consistent with the experimental one, indicating that the SC of LiSnSe$_2$ can be well described by the phonon-mediated mechanism.

Like enhancing SC of binary FeSe through external pressure, squeezing SnSe$_2$ can optimize the structure and carrier concentration, and then induce SC. In pressurized SnSe$_2$, the possibility of phase decomposition due to structural instability could soften the partial phonon modes, which enhances the electron-phonon coupling\cite{RN50}. In Li-intercalated SnSe$_2$, the newly-formed carrier layers can induce two kinds of effects. One is leading to phonon softening of SnSe$_2$, which could enhance the electron-phonon coupling and induce the SC according to the BCS theory. The other is that creating carrier layer of Li$^+$, which transfers electrons into SnSe$_2$ layer and realizes intrinsically heavily-doped SnSe$_2$\cite{RN26,RN27}. Our electric-field gating can increase carrier concentration by one or two orders of magnitude. The SnSe$_2$ with high carrier density has been demonstrated to be superconductors with \emph{T}$_c$= 3.9 K\cite{RN28} and 4.8 K\cite{RN29}. However, we noticed that this gating method may not change the compound with large bandgap ( \textgreater   2 eV) into metal, and the tuning ability is limited under the circumstances. Designing new devices, such as electric-double-layer transistor with liquid and solid ions, could further induce more superconductors from metal chalcogenides with large band gap.
\section{\uppercase\expandafter{\romannumeral4}.  CONCLUSION}
In summary, we report a semiconductor-superconductor transition of 4.8 K in Li-intercalated SnSe$_2$ through electric-field gating. A maximal \emph{T}$_c$ of 6.2 K is obtained by optimizing the gating time and chemical substitution of S for Se. More importantly, theoretical calculations reasonably predict the intercalated crystal structure and the metallic behaviors of LiSnSe$_2$. Besides, the partial phonon softening due to Li-intercalation strengthens the electron-phonon coupling, leading to the occurrence of SC. The combinations of experimental and theoretical results demonstrate that this method is highly effective for increasing carrier concentration and metallizing layered materials. Meanwhile, it is useful to study the intrinsic properties of superconducting layer because Li-intercalation of this method can keep the crystal structure intact to a larger extent.

\section{\uppercase\expandafter {ACKNOWLEDGMENTS}}

This work is financially supported by the National Natural Science Foundation of China (No. 51772322, 51532010, 11604290 and 51732010), by the National Key Research and Development Program of China (No. 2016YFA0300600, 2017YFA0304700), by the Chinese Academy of Sciences under Grant QYZDJ-SSW-SLH013.

\bibliography{reference}
\end{document}